# The Chaotic Art: Quantum Representation and Manipulation of Color

Guosheng Hu

China Academy of Art, Hangzhou, China
Email: huguosheng@caa.edu.cn
ORCID: 0000-0001-7844-7038

**Abstract:** Due to its unique computing principles, quantum computing technology will profoundly change the spectacle of color art. Focusing on experimental exploration of color qubit representation, color channel processing, and color image generation via quantum computing, this article proposes a new technical path for color computing in quantum computing environment, by which digital color is represented, operated, and measured in quantum bits, and then restored for classical computers as computing results. This method has been proved practicable as an artistic technique of color qubit representation and quantum computing via programming experiments in Qiskit and IBM Q. By building a bridge between classical chromatics and quantum graphics, quantum computers can be used for information visualization, image processing, and more color computing tasks. Furthermore, quantum computing can be expected to facilitate new color theories and artistic concepts.

Quantum computing is a new branch of information technology based on the quantum mechanics of subatomic particles. Unlike classical computing, which represents information by stringing binary bits encoded as 1s or 0s, quantum computing facilitates information representation and manipulation by harnessing quantum bits (qubits). Quantum computing represents data with qubits, which involves the superposition of a qubit state and entanglement of multi-qubits to introduce a new concept of computing process and enhance processing power [1]. This new computing technology will trigger revolutions, both in classical information technology and in art. Image processing and color computing are some of the fields that may undergo practical routine updates.

It is believed that quantum computing will inspire art exploration and engineering applications, because of its distinct functional differences from classical computing technology [2]. In classical computers, color is mainly simulated by mixing quantized channels. In the RGB color mode, for example, each color is mixed from channels of red, green, and blue [3]. According to the principle of digital bits, the value of each color channel is definite, while in quantum computing, the state of qubits is highly chaotic. The measurement value of a qubit lies in the probability value of |0> and |1> states, so the original digital principle of color is no longer applicable. Restricted by the current color display technology, quantum color computing still partly relies on digital color in classical computing. Therefore, it is necessary to find a method that merges quantum computing and classical computing, using the unique operating mode of the quantum computer to carry out color computing, and the results can be widely used in classical computers.

The project aims to explore the technical feasibility and artistic possibilities of color computing and graphic processing in quantum computing environment. This article proposes a method of qubit representation for color. Quantum computing programs were implemented for image processing validation, including artistic operations such as channel editing, entanglement of color and data, and entanglement of image and artistic elements. The originality of this project facilitates the interdisciplinary technology in the fields of graphic processing, quantum computing application, and CG art.

## QUBIT PRINCIPLES AND FEASIBILITY

Quantum computing uses qubits to represent information and build computing processes with quantum gates, which is different from the classical computing of binary system. First, a qubit is different from a binary bit, regarding the latter's chaotic superposition of $|0>$ and $|1>$ states. The qubit state is not identified until it collapses to 0 or 1 at measurement. However, this operational process is conducive toward searching for optimal solutions in a chaotic state, and that is why quantum computing has prominently efficient advantages over classical computation for some tasks (Fig 1). Second, the entanglement effect will appear in qubit superposition, which enhances computing power exponentially for certain algorithms, such as simulated annealing and artificial neural networks. Some researchers take advantage of quantum computing to solve problems that require larger computing tasks in classical computers. For example, Dahl proposed a quantum machine instruction (QMI) method for troubleshooting the four-color map problem [4].

Although the coloration technology of classical computers has progressed substantially, the attention on quantum computing is still far from satisfactory. The reason might be due to two factors: First, super-chaotic quantum color computing — which is contrary to the digital color of classical computing technology — is difficult for most colorists to accept at present; second, quantum computing technology itself is far from mature, and the corresponding color technology is yet to be developed. It is currently unclear what potential quantum computing has in the field of color.

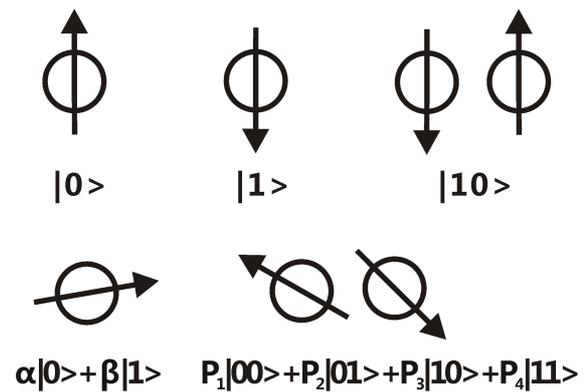

Fig 1. Principle of qubit superposition. The information capacity of qubits is far greater than that of the same number of digital (binary) bits, because each qubit is a chaotic superposition of $|0>$ and $|1>$.

However, researchers have tried image encryption with quantum computing methods for information security [5, 6]. They operate the color pixels of an image by recoding the image using qubits, processing and transmitting the image information, and then restoring the original image through the inverse operation, using binary based black-and-white dichotomy [7], a gray-scale channel method [8, 9], and the RGB based color of a multi-channel method [10]. The purpose of these operations is mainly protection, rather than art.

As quantum computing has boomed in recent years, colorists have begun to consider color computing according to the principles of quantum physics. As constructed by the channel values in classical computing, color can also be represented as qubit states for quantum computing. Bloch sphere, the geometric space representation of qubit states, can also be used to simulate the principle of color channels [11].

Due to the isomorphism of Bloch sphere structures and most color spaces, isomorphic mapping between them has been established to represent and manipulate color qubits. Nölle and Ömer proposed representing the three channels of color via a single qubit, because of the structural similarity of Bloch sphere and HSI color space [12]. This method comprises the separate Z-axis of Bloch sphere for intensity, the horizontal ring (equator) for hue, and the length of radius for saturation. Based on this, the authors asserted that a three-channel color could be represented via one single qubit that would make color computing much more concise and efficient. However, the representative point of the qubit state ($|\psi\rangle$) on quantum Bloch sphere only moves on the spherical surface, which means that one qubit can represent, at most, two channel values in unison, while the third channel can no longer be represented inside the sphere volume [13]. It might be inappropriate to represent three channels of a color within a single qubit. Therefore, I advocate that each color channel be represented by a qubit. Colors of single channels, such as grayscale, can be represented by a single qubit, while full color needs multi-qubit superposition.

**QUBIT COLOR REPRESENTATION**

To represent and manipulate color in quantum computing, qubits are loaded following the quantum mechanics in superposition and entanglement. The analogy of *Schrodinger's cat* for quantum mechanics indicates that a qubit is always in a chaotic superposition of $|0\rangle$ and $|1\rangle$ states before measurement [14]. Once observed, it will fall into either one of these two states. However, the superposed probabilities of $|0\rangle$ and $|1\rangle$ states embody the value of the qubit. To represent color via qubits, it is necessary to establish the mapping relationship between the color values and probability values of $|0\rangle$ or $|1\rangle$ states.

In quantum computing, a qubit is expressed as $|\psi\rangle = \alpha|0\rangle + \beta|1\rangle$, with $\alpha^2 + \beta^2 = 1$. This means that the superposition state lies in $\alpha$ or $\beta$, which are the probabilities of $|0\rangle$ or $|1\rangle$. To represent the superposition state in a geometric space, Bloch et al. introduced the Bloch sphere [15]. On the Bloch sphere, the trigonometry of the vertex depends on $\theta$ between 0 and $\pi$ with the diameter of the Z-axis as the bevel (Fig 2). To represent color in quantum computing, color values can be mapped to $\theta$ for qubit.

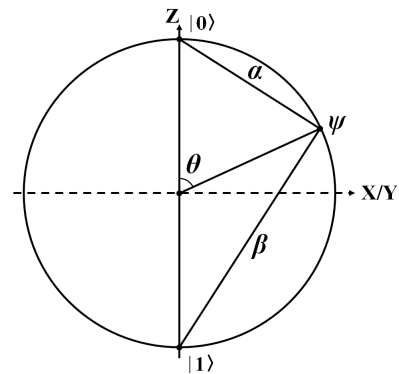

Fig 2. The relation between the angle of qubit ($\theta$) and the probability of two states $|0\rangle$ and $|1\rangle$. Where, $\psi$ is the quantum state, $\alpha$ and $\beta$ constitute the probabilities of the two states, respectively. Once the channel value of color is mapped to the $\psi$ angle ($\theta$) of the qubit, the color value can be transformed to the probability value of $|0\rangle$ ($\alpha$), then the color can be represented as a qubit state or multi-qubit states of combined channels.

**Representation of Single-Channel Color**

For single-channel colors, if the value (from 0 to channel maximum, usually 255) can be mapped to the $\psi$ angle $\theta$ (from 0 to $\pi$) on a qubit (see Fig. 2), then the probability value $\alpha$ is taken as the mapping value of the color as:

$$\theta = \text{acos}\left(2\frac{i}{max} - 1\right).$$

To set a qubit for color, the quantum gate is rotated to the corresponding angle. After the quantum measurement of the Z-axis, the $\alpha$ value of the result can be restored to color value as quantum circuit:

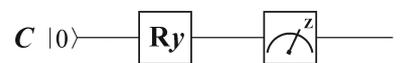

.

Thus, the color value is converted into the angle value of qubit, and the result can be restored to a new color value, according to the measured probability of the |0> state. Further operations can be programmed in between $\psi$ setting and the measurement.

In general, a quantum gate operation can be regarded as rotating the qubit vector of the Bloch sphere, so as to operate the superposition state of the qubit. When representing color, the channel value is first transformed into the Bloch angle, and the initial qubit state |0⟩ is rotated along the Y-axis via the R-gate. Then, operation tasks are programmed before the final measurement. The probability value $\alpha$ measured along the Z-axis of the qubit determines the resulting color.

**Representation of Multi-Channel Color**

Theoretically, R, G, and B channels in multi-channel color can be separately mapped to X-, Y-, and Z-axes on the Bloch sphere. They are represented as:

$$Z: R_0 |0>, R_1 |1>,$$

$$X: G_0 (|0>-|1>)/\sqrt{2}, G_1 (|0>+|1>)/\sqrt{2},$$

$$Y: B_0 (|0>-i|1>)/\sqrt{2}, B_1 (|0>+i|1>)/\sqrt{2};$$

or

$$R(Z): |0>, |1>,$$

$$G(X): |->, |+>,$$

$$B(Y): |-i>, |+i>.$$

Color in RGB mode can be described as a certain position on the Bloch sphere:

$$\hat{r} = |\psi\rangle = (x, y, z) = (\cos\varphi\sin\theta, \sin\varphi\theta, \cos\theta).$$

However, the three channels are independent of each other. The values of the three dimensions cannot move independently on the Bloch sphere, because any change in one channel value will lead to changes in the other two. It is the same dilemma as seen in Nölle and Ömer's method for HSI color mode. Besides, this method cannot deal with the problem of mapping grayscale colors of the uniform channel values. Therefore, each color is better represented by independent qubits for every channel, as |RGB>.

As long as RGB channel values from 0 to 255 are mapped to Bloch angles from 0 to $\pi$, the measurement value $\alpha$ of the |0> state is taken as the mapping value of the channel. Using three qubits for each RGB color, the sum of the probabilities (P values) of |0> states that superposed with other channels is the result value of that very channel. Thus, the state and probability relation of the qubits for R, G, and B channels of a color can be expressed as:

$$|\psi\rangle = P_1|R_0G_0B_0> + P_2|R_0G_0B_1> + P_3|R_0G_1B_0> + P_4|R_0G_1B_1> + P_5|R_1G_0B_0> + P_6|R_1G_0B_1> + P_7|R_1G_1B_0> + P_8|R_1G_1B_1>.$$

The value of the red channel is based on the probability of the items with the channel R in |0> state, as:

$$R = P_1 + P_2 + P_3 + P_4.$$

Therefore, when a color representation circuit composed of three qubits is measured, the P value of each channel should be:

$$R = P_{|000>} + P_{|001>} + P_{|010>} + P_{|011>};$$

$$G = P_{|000>} + P_{|100>} + P_{|001>} + P_{|101>};$$

$$B = P_{|000>} + P_{|110>} + P_{|100>} + P_{|010>}.$$

**Representation of Color Graphic**

An image can be considered a collection of color pixels, with each pixel described in the position coordinates and color channel values as (x, y, R, G, B). In quantum computing, each pixel can be represented as a unit of color parameters and pixel coordinates, as

$$|RGB>\otimes|xy> \text{ or } |RGBxy>.$$

Image representation methods such as FRQI (flexible representation for quantum images), MCQI (multi-channel representation for quantum images), NEQR (novel enhanced quantum representation of a digital image model), CQIR (Caraiman's quantum image representations), and so on [16], generally represent color channels and pixel coordinates in qubits, but a common problem is that each color requires a certain number of quantum bits. When a color in a RGB mode image is represented, for example, each color channel is represented by 8 qubits in the NCQI (novel color quantum images) method, and an image with x*y pixels occupies a large number of qubits, which certainly takes up far more computing capability, especially for large images [17]. It is critical to reduce the number of qubits required to represent the location of image pixels. Actually, two qubits are sufficient for the pixel information of a two-dimensional picture. In the measurements of |xy>, the following states are involved:

$$|\psi\rangle = P_1|x_0y_0\rangle + P_2|x_0y_1\rangle + P_3|x_1y_0\rangle + P_4|x_1y_1\rangle.$$

The coordinate value of a pixel corresponds to the probability sums that the measured item is |0>; that is,

$$x = P_1 + P_2;$$
$$y = P_3 + P_4.$$

The results are then mapped to the pixel coordinates to restore each pixel's position in the image, similar to that of the color channel values.

Another method is to transform the image pixel locations into a linear sequence for a single qubit. Each pixel is represented as |RGB$i$> ($i$ is the pixel index), and the pixel positions lie in the sequence of measured probability of |0> state. In this method, image size must be tagged, and the angles of $\psi$ are calculated according to the image's pixel sequence length.

## EXPERIMENTS IN COLOR OPERATION

To judge whether the color qubit representation method is feasible for quantum operation, two questions need to be settled. These include (1) whether the original image can be restored by reverse operation, which tests if the computing process is rigorous and the result controllable; and (2) whether this method has some advantages and characteristics different from classical computing, which assesses the value of quantum color computing. To answer the questions, image quantum operation experiments were carried out on *IBM Q* and the programming tool *Qiskit*.

**Operation and Restoration**

A key problem of quantum color computing is the super-chaotic qubit state; that is, whether the chaos leads to off-target, or aimless results out of control. It is an issue of whether quantum color computation is rigorous. A complete quantum program for an image consists of at least three steps: encoding, operating, and decoding. The uncertainty is supposed to be settled by operating the color images freely, and restoring the original color images through reverse operation. If the basic characteristics of the original image can be restored without overall changes after a series of quantum processes, it is proved that the operating process is regular and controllable.

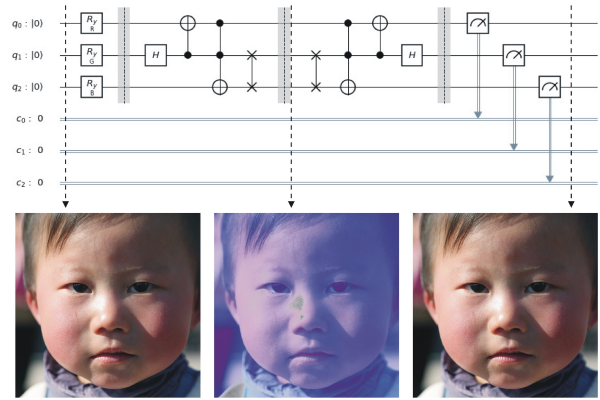

**Fig 3. A symmetrical quantum operation.** The left image is the original, the middle is the halfway effect after the half operation, and the right is the restored result after the flipped operation. The result shows that the original image can be essentially restored. It demonstrates that this method of color quantum computing is highly controllable and stable.

In fact, quantum gates, even the controlled gates for multi-qubit entanglement, such as CNOT (controlled-NOT), CCNOT (doubly controlled-NOT) and SWAP (swaps the state of two qubits), are reversible editing operations. When the combined gates of a quantum operation are symmetrically programmed, the manipulated objects will be restored, only being affected by qubit chaos, resulting in a certain extent of deviation. Here, two pictures were taken for an operating test, considering that there are various pixels of different colors. These experiments showed that the flipped operation could stably maintain the basic features of the original image (Fig 3 and Fig 4).

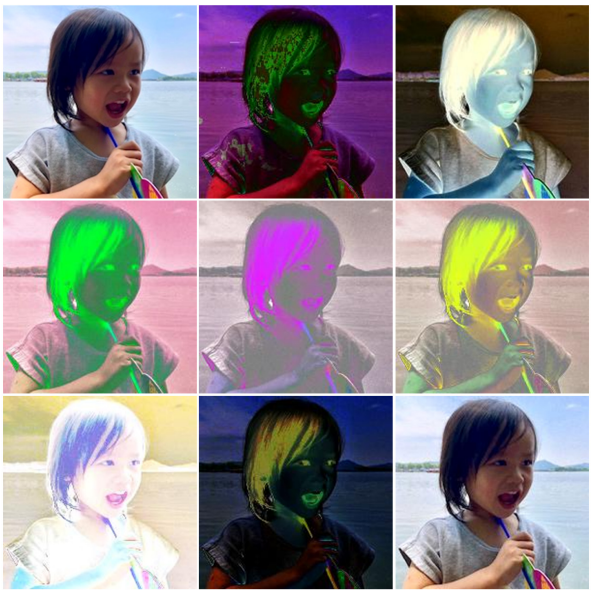

Fig 4. The restorable color effects of various quantum circuits. The top-left original picture was operated in various circuit steps with remarkable color changes, and finally restored to the bottom-right result after a symmetrically flipped operation.

**Editing of Color Elements**

Color elements are technical clues for evaluating color appearance, or variables for image processing. Operation on color elements can be seen in two forms: (1) manipulating channel relations of single colors or pixels to achieve new visual effects, and (2) controlling colors or images by channel variables to form interactive associations between objects. It is common to achieve color effects by entanglement among color channels in artistic image processing, modifying the intensities of, and relationships between channels, to change the characteristics of the image and render artistic effects.

Sometimes, free operation is achieved through channel conversion in different color modes. Fig 5 is a typical operation that represents the channel values of an image in RGB and HSV [18] color modes respectively, and edits the colors by entanglement among the channels.

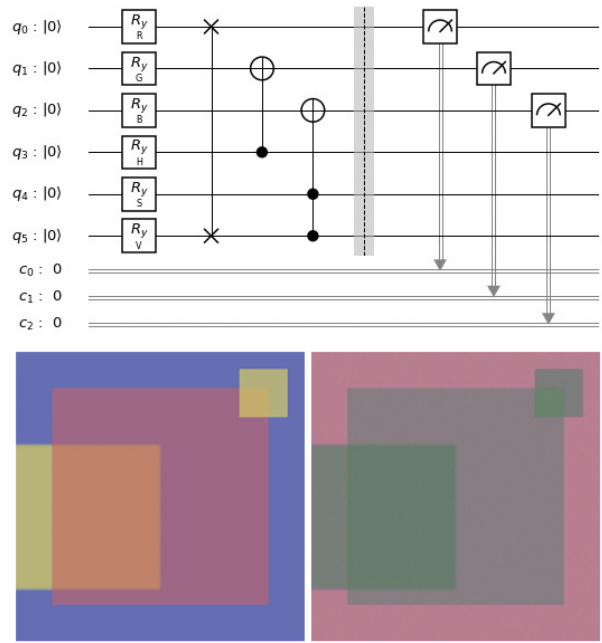

Fig 5. Entanglement operation between color channels of different modes. The top graph shows the operation process using qubits of HSV channels to replace or control the values of RGB channels. The bottom-left image is the original image, and the bottom-right is the edited outcome.

Entanglement between images can be operated for artistic style transfer, by replacing the color characteristics of an image with those of another. A new image can be achieved by color channel entanglement between images to produce style transfer, comprehensive mixing, structure mutation, and so on. Fig 6 shows an image that is the result of entanglement on the color relationships of the original images.

**Entanglement with External Data**

External data can be integrated into an image through entanglement, which can not only remold the color relations and visual effects, but also implant information as a watermark. This operation is useful for information security purposes. It can also be adopted as a medium of information visualization and interaction, because the external data introduced is usually continuously changing. Moreover, dynamic interaction effects of entanglement operations inspire new concepts of art; for example, the chaotic properties of quantum mechanics analogically trigger an association with synesthesia among various senses (Fig 7).

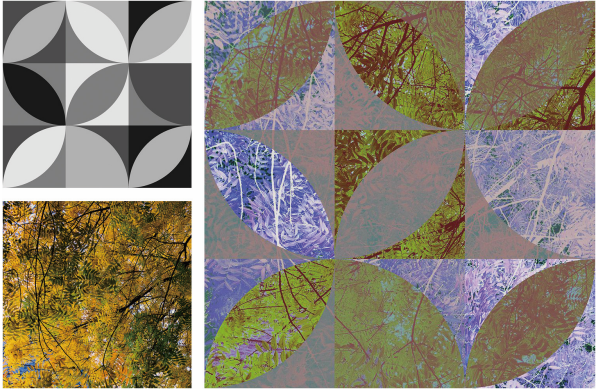

**Fig 6. A new image rendered by entanglement operation of two images. The entanglement operation combines the color channels of the images on the left. The right image is the generated image that embodies the information from the two originals.**

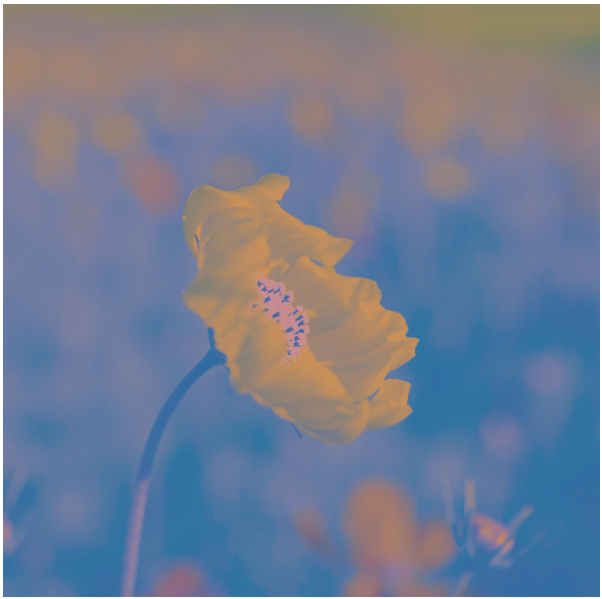

**Fig 7. Entanglement between a wildflower image and a musical chord in G major. The operation result retains the image details, while the color relation embodies the traits of the musical chord as is experienced in synesthesia. More details for musical color chords are available in the author's previous paper [19].**

### Operation for Image Pixel Position

It is necessary to represent and manipulate the position of each pixel in qubit for image processing. There are two methods of representing image pixel locations. One is to represent the pixel position in Cartesian coordinates as (x, y) corresponding to the image width and height axes. In this method, two qubits must be occupied for the abscissa and ordinate locations of the pixels. The other is to map the pixels of the image in a linear sequence set, where each pixel occupies a fixed position in the sequence, and only a single qubit is needed to represent the position of each pixel, with the size of the image being tagged.

Due to the chaotic trait of quantum computing, there will always be pixel dislocation after the quantum operation, and the dislocation is usually proportional to the size of the image. Fig 8 shows the results of manipulating the image pixels with two qubits, separately in coordinates x and y. The fuzzy noise in the resulting images was caused by the dislocation of the pixels, which deviated from their original positions due to qubit chaos. This phenomenon is even more remarkable in the single qubit representing method, because it produces an exponential number for pixel index.

Quantum operation always leads to unexpected effects due to the image pixel coordinate fluctuation caused by qubit chaos. However, it might be the very unexpected effects of quantum computing that enrich visual art and image processing with surprises, such as fuzzy, discrete, distortion, and other such effects.

### SIGNIFICANCE AND POTENTIAL

In the quantum operation of color elements, superposition and entanglement controlled by external data can lead to image processing and information visualization. The chaotic operation between color elements or images also inspires new ideas for color computing, color design and artistic purposes. While quantum computing brings uncertainty, it also implies more potential. As a visual medium, color usually involves the superposition and interaction of multiple factors, so that computational efficiency is naturally the main motivation for adopting quantum computing. However, artistic inspirations brought by this color representation and computation method could be more significant than those brought by a technical form per se. In fact, artistic originalities inspired by the relevant scientific theory and technology are constantly emerging. Although some artistic attempts probably do not involve quantum technology itself, the idea might be inspired by quantum mechanics or quantum computing. For example, Chavez and Conradi's artwork, Quantum Logos, was an attempt to use cultural archetypes or metaphors to explain quantum mechanics [20]. It allowed the audiences to experience an immersive, interactive, visual, and sonic access to approach quantum effects.

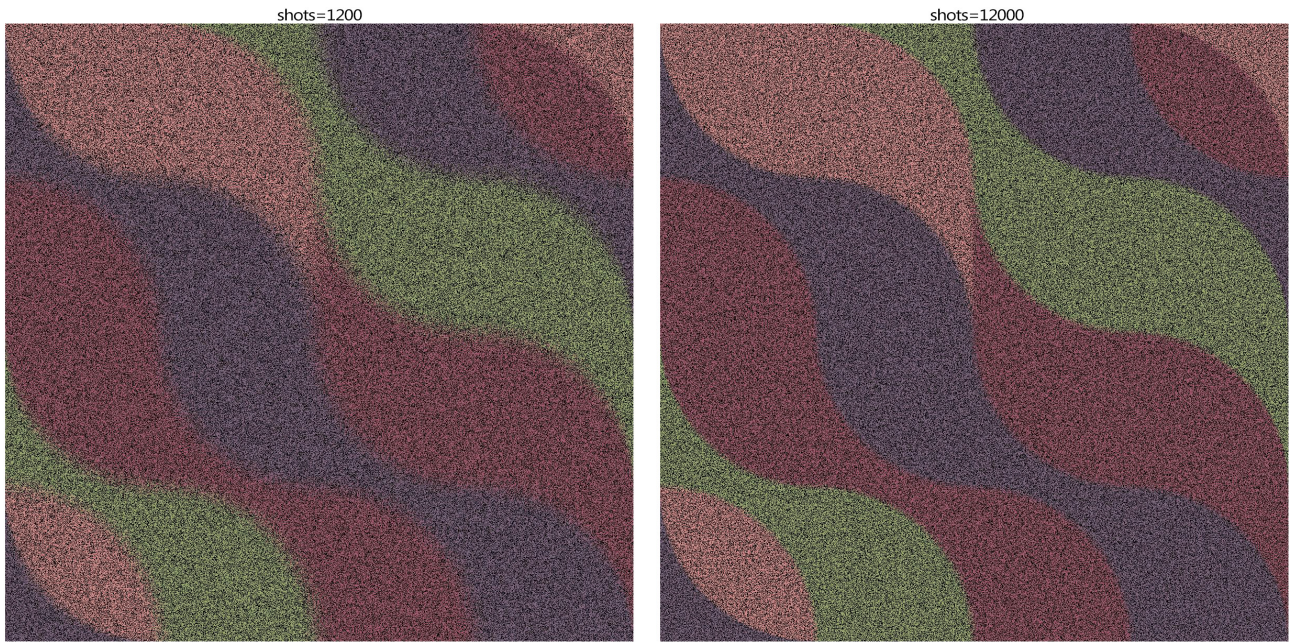

Fig 8. Dislocation of pixel position. Operation results of pixel coordinates fluctuate, forming image noise. This problem can be alleviated as the number of quantum measurements is increased. The measurements for the left image is 1,200 times, and the right 12,000 times.

Due to the limited qubits in a quantum computer at present, current programs for color image operation process the pixels in turn by loop operation, which occupies a few qubits each time. This limited operation does not benefit from the large-scale entanglement of a color image as an object of all the pixels superposed, without taking full advantage of quantum computing, such as quantum simulation, quantum annealing, etc. [21]. As the number of handling qubits in quantum computers increases to the point where all the pixels of an image can be processed simultaneously, quantum processing of color images will achieve a qualitative leap in processing efficiency and entanglement operation, and thus usher in a real computing revolution.

At present, the original data and output of color quantum computing still rely on the digital color of classical computing, where every color consists of red, green, and blue channels. With the development of new color simulation and computing technologies, color computing in the future may be completely different from today's color channel mixing. For example, a new color representation method might be found in the conceptual system of traditional Chinese color, in which every single color is a dynamically relative concept without specific quantitative principles, and color design is regarded as the contrast and harmonious relationship between various forces and elements. The elusive and interactive relationship among colors are exactly the same as the superposition and entanglement of qubits. This concept of color relations might be regarded as a valuable model in quantum computing. It presents a prospect worth pursuing in quantum computing technology, for a new theoretical color system.